\documentclass[a4paper,11pt]{article}

\usepackage{jcappub}
\usepackage[normalem]{ulem}

\usepackage{slashed}
\usepackage{bm}
\usepackage{latexsym,amssymb,amsmath,float,url,mathrsfs}
\usepackage{latexsym}
\usepackage{epstopdf}
\usepackage{amsfonts}
\usepackage{amsmath}
\usepackage{amssymb}
\usepackage{comment}
\usepackage{subfigure}
\usepackage{natbib}
\usepackage{hyperref}
\usepackage{pifont}
\usepackage{blindtext}
\usepackage{adjustbox}
\usepackage{multirow}
\usepackage{tabularx}
\usepackage[table]{xcolor}
\usepackage{orcidlink}
\usepackage{tikz}
\usetikzlibrary{tikzmark}
\usepackage{amssymb}
\usepackage{bbold}
\usepackage{blkarray}
\usepackage{graphicx}
\usepackage{amsfonts}
\usepackage{soul}
\usepackage{amssymb}
\usepackage{amsmath}
\usepackage{cancel}
\usepackage{tcolorbox}
\def\draftlabel#1{{\@bsphack\if@filesw {\let\thepage\relax
			\xdef\@gtempa{\write\@auxout{\string
					\newlabel{#1}{{\@currentlabel}{\thepage}}}}}\@gtempa
		\if@nobreak \ifvmode\nobreak\fi\fi\fi\@esphack}
	\gdef\@eqnlabel{#1}}
\def\@eqnlabel{}
\def\@vacuum{}
\def\draftmarginnote#1{\marginpar{\raggedright\scriptsize\tt#1}}

\def\draft{\oddsidemargin -.5truein
		\def\@oddfoot{\sl preliminary draft \hfil
			\rm\thepage\hfil\sl\today\quad\militarytime}
		\let\@evenfoot\@oddfoot \overfullrule 3pt
		\let\label=\draftlabel
		\let\marginnote=\draftmarginnote
		\def\@eqnnum{(\theequation)\rlap{\kern\marginparsep\tt\@eqnlabel}%
			\global\let\@eqnlabel\@vacuum}  }
	
	\def\preprint{\twocolumn\sloppy\flushbottom\parindent 2em
		\leftmargini 2em\leftmarginv .5em\leftmarginvi .5em
		\oddsidemargin -.5in    \evensidemargin -.5in
		\columnsep .4in \footheight 0pt
		\textwidth 10.in        \topmargin  -.4in
		\headheight 12pt \topskip .4in
		\textheight 6.9in \footskip 0pt
		\def\@oddhead{\thepage\hfil\addtocounter{page}{1}\thepage}
		\let\@evenhead\@oddhead \def\@oddfoot{} \def\@evenfoot{} }
	
	\def\numberbysection{\@addtoreset{equation}{section}
		\def\theequation{\thesection.\arabic{equation}}}
	
	\def\underline#1{\relax\ifmmode\@@underline#1\else
		$\@@underline{\hbox{#1}}$\relax\fi}

	\def\titlepage{\@restonecolfalse\if@twocolumn\@restonecoltrue\onecolumn
		\else \newpage \fi \thispagestyle{empty}\c@page\z@
		\def\thefootnote{\fnsymbol{footnote}} }
	
	\def\endtitlepage{\if@restonecol\twocolumn \else \newpage \fi
		\def\thefootnote{\arabic{footnote}}
		\setcounter{footnote}{0}}  

\catcode`@=12
\relax

\def\figcap{\section*{Figure Captions\markboth
		{FIGURECAPTIONS}{FIGURECAPTIONS}}\list
	{Figure \arabic{enumi}:\hfill}{\settowidth\labelwidth{Figure
			999:}
		\leftmargin\labelwidth
		\advance\leftmargin\labelsep\usecounter{enumi}}}
 \relax
\def\tablecap{\section*{Table Captions\markboth
		{TABLECAPTIONS}{TABLECAPTIONS}}\list
	{Table \arabic{enumi}:\hfill}{\settowidth\labelwidth{Table
			999:}
		\leftmargin\labelwidth
		\advance\leftmargin\labelsep\usecounter{enumi}}}
 \relax
\def\reflist{\section*{References\markboth
		{REFLIST}{REFLIST}}\list
	{[\arabic{enumi}]\hfill}{\settowidth\labelwidth{[999]}
		\leftmargin\labelwidth
		\advance\leftmargin\labelsep\usecounter{enumi}}}
 \relax

	\makeatletter
	\newcounter{pubctr}
	\def\publist{\@ifnextchar[{\@publist}{\@@publist}}
	\def\@publist[#1]{\list
		{[\arabic{pubctr}]\hfill}{\settowidth\labelwidth{[999]}
			\leftmargin\labelwidth
			\advance\leftmargin\labelsep
			\@nmbrlisttrue\def\@listctr{pubctr}
			\setcounter{pubctr}{#1}\addtocounter{pubctr}{-1}}}
	\def\@@publist{\list
		{[\arabic{pubctr}]\hfill}{\settowidth\labelwidth{[999]}
			\leftmargin\labelwidth
			\advance\leftmargin\labelsep
			\@nmbrlisttrue\def\@listctr{pubctr}}}
	 \relax
	\makeatother
	\newskip\humongous \humongous=0pt plus 1000pt minus 1000pt

	\newif\ifdtup

	\relax
	
	\def\be{\begin{equation}}
		\def\ee{\end{equation}}
	\def\ba{\begin{eqnarray}}
		\def\ea{\end{eqnarray}}

\newcommand{\dd}{\mathrm{d}}
\newcommand{\ii}{\mathrm{i}}
\newcommand{\re}{\mathrm{e}}
\newcommand{\rs}{r_s}
\newcommand{\rstar}{r_*}

\newcommand{\Pc}{P_C}
\newcommand{\Vref}{V^{\rm ref}}
\newcommand{\Lref}{L^{\rm ref}}

\newcommand{\abs}[1]{\left|#1\right|}
\newcommand{\paren}[1]{\left(#1\right)}

	\def \black hole { {\bar h} }

	\def\IR{\relax{\rm I\kern-.18em R}}
	\def\IL{\relax{\rm I\kern-.18em L}}
	
	\def\inv{^{\raise.15ex\hbox{${\scriptscriptstyle -}$}\kern-.05em 1}}

	\def\bea{\begin{eqnarray}}
		\def\eea{\end{eqnarray}}

	\definecolor{markcolor2}{rgb}{1,0,0}
	
	\definecolor{markcolor3}{rgb}{0,1,0}

    \def\dd{{\rm d}}

\usepackage{graphics,appendix,afterpage,makecell} 

\def\abs#1{\left| #1\right|}

\definecolor{oucrimsonred}{rgb}{0.6, 0.0, 0.0}
\definecolor{persianblue}{rgb}{0.11, 0.22, 0.73}
\definecolor{forestgreen}{rgb}{0.13,0.35,0.13}
\definecolor{lightgray}{rgb}{0.83, 0.83, 0.83}
\definecolor{cornellred}{rgb}{0.7, 0.11, 0.11}
\definecolor{navyblue}{rgb}{0.0, 0.0, 0.5}
\definecolor{amethyst}{rgb}{0.6, 0.4, 0.8}
\definecolor{yellow}{rgb}{1.0, 1.0, 0.0}
\definecolor{firebrick}{rgb}{0.7, 0.13, 0.13}
\definecolor{tangerineyellow}{rgb}{1.0, 0.8, 0.0}
\definecolor{deepfuchsia}{rgb}{0.76, 0.33, 0.76}
\definecolor{amber}{rgb}{1.0, 0.75, 0.0}
\definecolor{VioletRed4}{rgb}{0.55, 0.13, .32}
\definecolor{indiagreen}{rgb}{0.07, 0.53, 0.03}
\definecolor{VioletRed4}{rgb}{0.55, 0.13, .32}

\definecolor{oucrimsonred}{rgb}{0.6, 0.0, 0.0}
\newcommand\vertarrowbox[3][6ex]{%
  \begin{array}[t]{@{}c@{}} #2 \\
  \left\uparrow\vcenter{\hrule height #1}\right.\kern-\nulldelimiterspace\\
  \makebox[0pt]{\scriptsize#3}
  \end{array}%
}
\hypersetup{
     colorlinks   = true,
     citecolor    = violet,
     urlcolor     = violet,
     linkcolor    = violet}

\definecolor{verdechiaro}{rgb}{0.6,1,0.6}
\definecolor{giallochiaro}{rgb}{1,1,0.6}
\definecolor{bluscuro}{rgb}{0.15, 0.2, 0.9}
\definecolor{verdes}{rgb}{0.1, 0.5, 0.1}%
\definecolor{tangerineyellow}{rgb}{1.0, 0.8, 0.0}

\definecolor{americanrose}{rgb}{1.0, 0.01, 0.24}
\definecolor{cobalt}{rgb}{0.0, 0.28, 0.67}
\definecolor{brandeisblue}{rgb}{0.0, 0.44, 1.0}
\definecolor{mycolor}{rgb}{0.0, 0.0, 0.5}
\definecolor{oxfordblue}{rgb}{0.0, 0.13, 0.28}
\definecolor{azure}{rgb}{0.0, 0.5, 1.0}
\definecolor{turquoiseblue}{rgb}{0.0, 1.0, 0.94}
\newtcolorbox{mynewbox}[1]{colback=white!5!white,colframe=azure!75!black,fonttitle=\bfseries,title=#1}
\newtcolorbox{mybox}{colback=mycolor!5!white,colframe=azure!75!black}
\newtcolorbox{mynamedbox}[1]{colback=mycolor!5!white,colframe=azure!75!black,title=#1}
\definecolor{venetianred}{rgb}{0.78, 0.03, 0.08}
\newtcolorbox{mynamedbox1}[1]{colback=venetianred!5!white,colframe=venetianred!80!black,title=#1}
\newtcolorbox{mynamedbox2}[1]{colback=azure!5!white,colframe=azure!80!black,title=#1}

\definecolor{verdes}{rgb}{0.1, 0.5, 0.1}%
\definecolor{cornellred}{rgb}{0.7, 0.11, 0.11}

\definecolor{VioletRed4}{rgb}{0.55, 0.13, .32}

\definecolor{rossocorsa}{rgb}{0.83, 0.0, 0.0}

\title{Radial Mirror Scattering and the QNM Convergence Region }

\author[a]{Alex~Kehagias\orcidlink{0000-0001-6080-6215},}
\author[b]{Antonio~Riotto\orcidlink{0000-0001-6948-0856}}

\affiliation[a]{Physics Division, National Technical University of Athens, Athens 15780, Greece}

\affiliation[b]{Department of Theoretical Physics and Gravitational Wave Science Center,  \\
24 quai E. Ansermet, CH-1211 Geneva 4, Switzerland}
\affiliation[c]{INFN, Sezione di Roma, Piazzale Aldo Moro 2, 00185 Rome, Italy}

\abstract{We revisit the convergence region of the quasinormal modes expansion of Schwarzschild retarded Green functions from a radial scattering viewpoint. The tortoise coordinate admits a natural reflection about a distinguished point, which maps the original Regge-Wheeler problem to a mirror radial problem with the same quasinormal mode spectrum. Although this reflection is not a spacetime symmetry and does not leave the potential invariant, it gives a simple image interpretation of the second lightcone distance that controls convergence. Equivalently, after folding the radial line at the reflection point, the direct and mirror contributions arise as diagonal and off-diagonal propagation channels of a two-component half-line problem. We also relate this structure to the AdS$_2$ Green function, where the same direct-plus-image lightcone structure arises from a genuine boundary-bouncing null geodesic. This provides a spectral interpretation of the convergence condition and clarifies the role of the reflection point in the Schwarzschild radial Green function. 
}

\emailAdd{kehagias@central.ntua.gr}
\emailAdd{Antonio.Riotto@unige.ch}
\begin{document}
\maketitle
\flushbottom


\section{Motivation and main idea}

The quasinormal mode (QNM) part of the retarded Green function for Schwarzschild perturbations has the expansion
\begin{equation}
        G_R^{\rm QNM}(t,t';r,r')
        =\sum_{n=0}^{\infty} g_n(r,r')\,
        \re^{-\ii\omega_n(t-t')} .
        \label{eq:qnm-sum}
\end{equation}
The coefficients $g_n(r,r')$ are the residues of the frequency-domain Green function at the QNM poles, and $\omega_n$ are the QNM frequencies.  The convergence of this sum is a important question because the high-overtone terms are not uniformly small for all real times.  The large-overtone frequencies are approximately equally spaced along the negative imaginary axis so that  the QNM sum behaves like a power series in a complex exponential of time, and therefore, its radius of convergence is controlled by the closest singularity of the analytically continued Green function.

For the Schwarzschild Green function, the convergence region is not determined only by the ordinary direct lightcone between the observation point and the source point.  In terms of the tortoise coordinate $\rstar$, the convergence condition takes the form \cite{Arnaudo:2025kit,Arnaudo:2025uos,ArnaudoWithers2026,DeAmicis:2026wqd,Kuntz:2026xep}
\begin{equation}
        t-t'
        >
        \max\left\{
        \abs{\rstar-\rstar'},
        \abs{\rstar+\rstar'-2C}
        \right\} .
        \label{eq:conv-reg}
\end{equation}
The first term is immediately recognized as the direct radial lightcone distance in the two-dimensional $(t,\rstar)$ sector.  The second term is more surprising as it  contains the sum of the two tortoise coordinates.  This suggests an image construction, because the expression can be rewritten as the distance between $\rstar$ and the reflected position of the source point.

The additive constant $C$ in Eq. \eqref{eq:conv-reg} is arbitrary and can be put to zero, but we keep it  as bookkeeping parameter.  Let us define the radial reflection operator $\Pc$ by
\begin{equation}
        \Pc(\rstar)=2C-\rstar .
        \label{eq:reflection-intro}
\end{equation}
The fixed point of this reflection is $\rstar=C$.  If the original source is located at \(r_*'\), its reflected image is
\begin{equation}
r_{*,{\rm im}}'=P_C(r_*')=2C-r_*' .
\end{equation}
The distance from the observation point $\rstar$ to this image source  is
\begin{equation}
        \abs{\rstar-r_{*,{\rm im}}'}
        =\abs{\rstar+\rstar'-2C} .
        \label{eq:image-distance-intro}
\end{equation}
Thus the second term in the convergence condition \eqref{eq:conv-reg} is the ordinary radial lightcone distance from the image of the source.  

The important point is that this reflection is not a true symmetry of the full Schwarzschild spacetime.  It is also not a symmetry of the RW potential itself, because in general
\begin{equation}
        V(2C-\rstar)\neq V(\rstar) .
        \label{eq:noninvariant-intro}
\end{equation}
Instead, the reflection maps the original radial problem to a reflected radial problem.  The reflected problem has potential $V^{\rm ref}(\rstar)=V(2C-\rstar)$.  The two one-dimensional radial operators are related by conjugation with the reflection operator.  Consequently, they are isospectral.  The QNM frequencies of the reflected potential are the same as those of the original potential, provided the boundary conditions are reflected at the same time.

This isospectrality does not by itself prove the convergence region of the original  Green function.  The convergence of the QNM sum depends not only on the positions of the poles, but also on the large-overtone residues.  What reflection does is to  identify the image optical length, and then the residue asymptotics to tells us that this image optical length actually appears in the original Green function.  More precisely, the large-overtone residues contain phases of the form
\begin{equation}
        \exp\left[\ii\omega_n(\rstar+\rstar'-2C)\right] .
        \label{eq:image-phase-intro}
\end{equation}
When this phase is combined with the time dependence $\exp[-\ii\omega_n(t-t')]$ and with the large-overtone spacing of the QNM frequencies, the QNM sum becomes a logarithmic series.  The singularity of this logarithmic series gives the reflected or bounced lightcone.  The direct lightcone is obtained in the same way from the ordinary direct phases involving $\rstar-\rstar'$.

The aim of the reflection viewpoint is therefore not to replace the complex-time bouncing-singularity analysis \cite{ArnaudoWithers2026}, but   rather to give a complementary radial scattering interpretation, where  the bounce radius is the fixed point of the radial reflection.  The reflected lightcone is the direct lightcone from the reflected source.  The reflected radial problem is isospectral to the original one.  Finally, the high-overtone residues provide the dynamical input which turns this mirror optical length into a singularity of the analytically continued Green function.

The paper is organized as follows. In section 2 we discuss the RW equation in tortoise coordinates, while the frequency domain Green function and mirror lightcones are discussed in section 3. The origin of the mirror phase is presented in section 4 as well as the AdS$_2$ link in section 5. Our conclusions are in section 6.

\section{The Regge--Wheeler equation in tortoise coordinates}

The background of a spinless black hole is described by the Schwarzschild metric, which is written as
\begin{equation}
        \dd s^2
        =-f(r)\,\dd t^2+\frac{\dd r^2}{f(r)}+r^2\dd\Omega_2^2,
        \qquad
        f(r)=1-\frac{\rs}{r},
        \qquad
        \rs=2M .
        \label{eq:schwarzschild}
\end{equation}
In the following, we wil mainly work with the  tortoise coordinate, which is defined as 
\begin{equation}
        \rstar
        =r+\rs\log\paren{\frac{r}{\rs}-1}+C .
        \label{eq:tortoise}
\end{equation}
where the constant $C$ is arbitrary.  Different choices of $C$ simply translate the origin of $\rstar$.  However, keeping $C$ explicit determines the fixed point of the reflection $P_C$ to be at   $\rstar=C$.  In terms of $\rstar$, the radial coordinate is expressed outside of th horizon as 
\begin{equation}
        r(\rstar)
        =\rs\left[1+W_0\left(\re^{(\rstar-C)/\rs-1}\right)\right] ,
        \label{eq:inv-tor}
\end{equation}
where  $W_0$ is the principal branch of the Lambert function. Then, the fixed point of the reflection $P_C$, also called bounce radius, is   in terms of the radial coordinate $r$
\cite{ArnaudoWithers2026} 
\begin{equation}
        r_{\rm bounce}
        =\rs\left[1+W_0(e^{-1})\right] .
        \label{eq:rbounce}
\end{equation}
The bounce radius is the radius reached by the limiting complex null geodesic that starts from the exterior, reflects off the Schwarzschild singularity $r=0$, and returns to the exterior. In tortoise coordinates this reflected trajectory appears to bounce at $\rstar^{\rm bounce}=C$.

\begin{figure}[htbp]
  \centering
\includegraphics[width=0.5\textwidth]{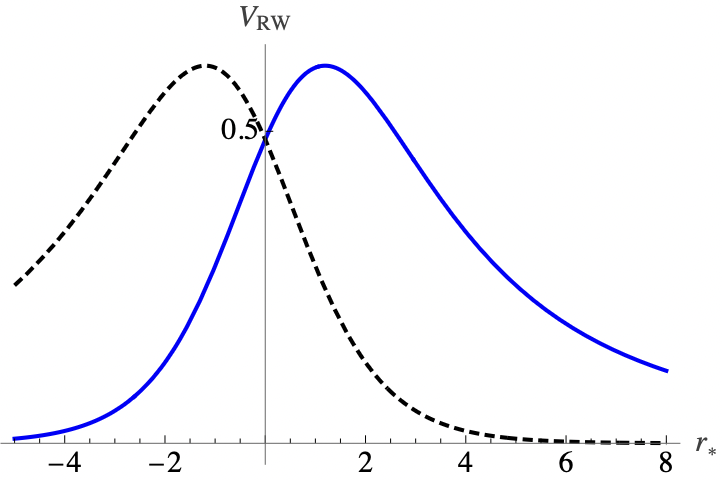}
  \caption{The potential $V_{\rm RM}$ (solid blue curve) and the reflected one $V_{\rm RW}^{\rm ref}$ (dashed black curve)  as a function of $\rstar$ 
           in units of $2M$ and $C=0$.}
  \label{fig:VRM}
\end{figure}

The radial wave equation in the Regge--Wheeler (RW) gauge for a massless perturbation of spin $s$, after the usual separation into spherical harmonics and after the standard radial rescaling,  can be written as
\begin{equation}
        \frac{\dd^2\phi}{\dd\rstar^2}
        +\left[\omega^2-V_s(r)\right]\phi=0 ,
        \label{eq:rw-equation}
\end{equation}
where the spin-dependent Schwarzschild potential is
\begin{equation}
        V_s(r)
        =f(r)\left[
        \frac{\ell(\ell+1)}{r^2}
        +(1-s^2)\frac{\rs}{r^3}
        \right] .
        \label{eq:spin-s-potential}
\end{equation}
For axial gravitational perturbations we will discuss below, we have $s=2$, so that the RW potential
turns out to be
\begin{equation}
        V_{\rm RW}(r)
        =f(r)\left[
        \frac{\ell(\ell+1)}{r^2}
        -\frac{3\rs}{r^3}
        \right] .
        \label{eq:rw-potential-r}
\end{equation}
It is convenient to write the $V_{\rm RW}$ potential  as a function of $\rstar$ by using Eq. \eqref{eq:inv-tor}.  Let us define 
\begin{equation}
w(\rstar)=W_0\left(\re^{(\rstar-C)/\rs-1}\right) 
,    \label{eq:w-def}
\end{equation}
so that  the gravitational RW potential becomes
\begin{equation}
        V_{\rm RW}(\rstar)
=\frac{w(\rstar)\left[\ell(\ell+1)(1+w(\rstar))-3\right]}
{\rs^2(1+w(\rstar))^4} ,
        \label{eq:rw-potential-w}
\end{equation}
or, for general spin $s$
\begin{equation}
        V_s(\rstar)
        =\frac{w(\rstar)}{\rs^2(1+w(\rstar))^4}
        \left[
        \ell(\ell+1)(1+w(\rstar))+(1-s^2)
        \right] .
        \label{eq:spin-s-potential-w}
\end{equation}
The reflected potential $V_{\rm RW}^{\rm ref}$ is obtained by replacing $\rstar$ by $2C-\rstar$ in $V_{\rm RW}$, which, if 
\begin{equation}
        w_{\rm ref}(\rstar)=w(2C-\rstar)
        =W_0\left(\re^{-(\rstar-C)/\rs-1}\right) ,
        \label{eq:w-reflected}
\end{equation}
turns out to be 
\begin{equation}
        V_{\rm RW}^{\rm ref}(\rstar)
        =V_{\rm RW}(2C-\rstar)
        =\frac{w_{\rm ref}(\rstar)
        \left[\ell(\ell+1)(1+w_{\rm ref}(\rstar))-3\right]}
        {\rs^2(1+w_{\rm ref}(\rstar))^4} .
        \label{eq:rw-ref-potential}
\end{equation}
  The original potential and the reflected potential are mirror images around $\rstar=C$, but the original potential is not symmetric around that point, see Fig. \ref{fig:VRM}.

\subsection{Reflected radial operators and exact isospectrality}

Let us now consider the  one-dimensional radial differential operator
\begin{equation}
        L=-\frac{\dd^2}{\dd\rstar^2}+V(\rstar),
        \label{eq:L-original}
\end{equation}
where the radial mode equation may be written as
\begin{equation}
        L\psi=\omega^2\psi .
        \label{eq:eigenvalue-equation}
\end{equation}
Then, the reflection operator \eqref{eq:reflection-intro} acts on the modes $\psi(\rstar)$ as
\begin{equation}
        (\Pc\psi)(\rstar)=\psi_{\rm ref}(\rstar)=\psi(2C-\rstar) ,
        \label{eq:Pc-def}
\end{equation}
and the reflected operator $\Lref$  is
\begin{equation}
        \Lref=-\frac{\dd^2}{\dd\rstar^2}+\Vref(\rstar) .
        \label{eq:L-ref}
\end{equation}
Now the original and reflected operators are related by
\begin{equation}
        \Lref=\Pc L\Pc^{-1} ,
        \label{eq:conjugation}
\end{equation}
which leads to their  isospectrality.  The second derivative is unchanged by reflection, because applying the reflection before and after differentiating twice leaves the operator $-\dd^2/\dd\rstar^2$ invariant.  The potential term is changed into multiplication by $V(2C-\rstar)$.  Therefore the reflected operator is exactly the conjugate of the original operator.

If $\psi(\rstar)$ solves the original radial equation with frequency $\omega$, then $\Pc\psi$ solves the reflected radial equation with the same frequency.  Indeed, we find 
\begin{equation}
        \Lref(\Pc\psi)=\omega_{\rm ref}^2 \Pc\psi=\Pc L\psi=\omega^2(\Pc\psi) ,
        \label{eq:eigenfunction-map}
\end{equation}
and thus, the two operators $L$ and $\Lref$ have the same eigenvalues or resonance frequencies ($\omega_{\rm ref}=\omega$), provided one reflects the boundary conditions at the same time.

It is straightforward to determine the boundary conditions of the reflected operator $\Lref$ if we employ   the centered coordinate 
\begin{eqnarray}
    x=\rstar-C, \label{eq:x-rstar}
\end{eqnarray}
 so that 
\begin{equation}
        L=-\partial_x^2+V(x),
        \qquad
        \Lref=-\partial_x^2+V(-x)
        \label{eq:centered-L}. 
\end{equation}
We will now  verify below that the QNM boundary conditions are preserved by the reflection. In the original problem, a QNM satisfies ingoing boundary condition at the horizon, 
\begin{equation}
\psi(x)\sim e^{-\ii\omega x},
\qquad x\to-\infty ,
\end{equation}
and outgoing boundary condition at spatial infinity
\begin{equation}
\psi(x)\sim e^{+\ii\omega x},
\qquad x\to+\infty .
\end{equation}
Now consider the reflected wavefunction
\begin{equation}
\psi_{\rm ref}(x)=\psi(-x).
\end{equation}
When $x\to-\infty$, the argument $-x$ goes to $+\infty$, so the reflected wavefunction captures the right-end behavior of the original solution and becomes
\begin{equation}
\psi_{\rm ref}(x)=\psi(-x)\sim e^{+\ii\omega(-x)}=e^{-\ii\omega x}.
\end{equation}
This is precisely the correct ingoing condition at the left end of the reflected problem. Similarly, when $x\to+\infty$, the argument $-x$ goes to $-\infty$, so the reflected wavefunction captures the left-end behavior of the original solution and becomes
\begin{equation}
\psi_{\rm ref}(x)=\psi(-x)\sim e^{-\ii\omega(-x)}=e^{+\ii\omega x}.
\end{equation}
This is the correct outgoing condition at the right end. Thus the reflection exchanges the two asymptotic ends, but the sign change in the argument compensates this exchange. The left and right scattering channels are interchanged, while the QNM boundary-value problem is mapped into the corresponding QNM boundary-value problem for the reflected potential. In other words, 
the reflected function $\psi_{\rm ref}(x)=\psi(-x)$ satisfies the corresponding outgoing conditions for the reflected potential $V(-x)$.
We emphasize that this equality is a statement about two different scattering problems, the original one and the reflected one.  It is not a statement that the original Schwarzschild potential is invariant under the reflection.

Notice that for scattering amplitudes, the reflection exchanges the two asymptotic ends.  The transmission amplitude is unchanged, while the reflection amplitude for incidence from the left is exchanged with the reflection amplitude for incidence from the right.  If the original transmission amplitude is $T(\omega)$ and the original left and right reflection amplitudes are $R_L(\omega)$ and $R_R(\omega)$, then the reflected problem has
\begin{equation}
        T_{\rm ref}(\omega)=T(\omega),
        \qquad
        R_{L,\rm ref}(\omega)=R_R(\omega),
        \qquad
        R_{R,\rm ref}(\omega)=R_L(\omega) .
        \label{eq:scattering-data}
\end{equation}
The poles of the scattering matrix are therefore unchanged since the transmission amplitude is invariant under reflection, although the left and right interpretation of the scattering data is exchanged.

\subsection{The radial Kruskal involution}

The radial reflection has a natural expression in Kruskal--Szekeres coordinates.
In the right exterior region we use
\begin{equation}
        U=-\exp\left[-\frac{t-\rstar}{2\rs}\right],
        \qquad
        V=\exp\left[\frac{t+\rstar}{2\rs}\right] ,
        \label{eq:kruskal}
\end{equation}
so that we have
\begin{equation}
        UV=-\re^{\rstar/\rs} .
        \label{eq:UV-rstar}
\end{equation}
Because the tortoise coordinate contains the additive constant $C$, the analytically continued Schwarzschild singularity $r=0$ lies at
$ UV=+\re^{C/\rs}$.
The reflected image of this locus in the radial Kruskal plane is
\begin{equation}
        UV=-\re^{C/\rs} ,
        \label{eq:bounce-locus}
\end{equation}
which by using \eqref{eq:UV-rstar}, it corresponds  to $\rstar=C$. Therefore,  the bounce radius is represented in Kruskal coordinates by the curve $UV=-\re^{C/\rs}$.

Let us now define a radial Kruskal mirror map by
\begin{equation}
  (U,V)\quad \longrightarrow\quad (U_{\rm ref},V_{\rm ref})=\left(     -\frac{\re^{C/\rs}}{V}, -\frac{\re^{C/\rs}}{U}\right). 
        \label{eq:pC-kruskal}
\end{equation}
For $C=0$ this is simply the inversion
\begin{equation}
       (U,V)\quad \longrightarrow\quad \left(-\frac{1}{V},-\frac{1}{U}\right) .
        \label{eq:p0-kruskal}
\end{equation}
Applying this map twice gives back the original point, so it is an involution of the two-dimensional radial Kruskal plane. In terms  of the coordinates $(t,\rstar)$  we have
\begin{equation}
    (t,\rstar)\quad \longrightarrow\quad   (t_{\rm ref},\rstar^{\rm ref})=(t,2C-\rstar),
        \label{eq:pC-action}
\end{equation}
so that the fixed locus is at $\rstar^{\rm ref}=\rstar$, i.e., at $\rstar=C$.  Hence, in Kruskal coordinates,the bounce radius is the fixed curve $UV=-\re^{C/\rs}$ of the two-dimensional radial Kruskal involution \eqref{eq:pC-action}.
Clearly this involution is not a four-dimensional Schwarzschild isometry.

\section{Frequency-domain Green functions and mirror lightcones}

Let us denote $\phi^{\rm in}_{\omega}$  the solution which is purely ingoing at the horizon, and  $\phi^{\rm up}_{\omega}$  the solution which is purely outgoing at infinity, with   asymptotic behaviour
\begin{equation}
\phi^{\rm in}_{\omega}\sim \re^{-\ii\omega\rstar}
\quad \hbox{near the horizon},
\qquad
\phi^{\rm up}_{\omega}\sim \re^{+\ii\omega\rstar}
\quad \hbox{near infinity} .
\label{eq1}
\end{equation}
Then, the  frequency-domain Green function $\widetilde G_{\omega}$  for the original radial operator
which satisfies
\begin{equation}
\left[\frac{\dd^2}{\dd\rstar^2}+\omega^2-V(\rstar)\right]
\widetilde G_{\omega}(\rstar,\rstar')
=
\delta(\rstar-\rstar') .
\label{eq}
\end{equation} 
can be written as 
\begin{eqnarray}
    \widetilde G_{\omega}(\rstar,\rstar') = \begin{cases} \dfrac{\phi^{\rm in}_{\omega}(\rstar)\phi^{\rm up}_{\omega}(\rstar')}{W(\omega)} & \quad \text{if   }\quad \rstar<\rstar' \\ \\ \dfrac{\phi^{\rm in}_{\omega}(\rstar')\phi^{\rm up}_{\omega}(\rstar)}{W(\omega)} 
\label{eq2} & \quad \text{if  } \quad\rstar>\rstar' ,\end{cases}
\end{eqnarray}
where
\begin{equation}
W(\omega)
=
\phi^{\rm up}_{\omega}\partial_{\rstar}\phi^{\rm in}_{\omega}
-
\phi^{\rm in}_{\omega}\partial_{\rstar}\phi^{\rm up}_{\omega},
\label{eq3}
\end{equation}
is the Wronskian. The QNM frequencies are then the zeros of $W(\omega)$ after the appropriate analytic continuation and outgoing boundary conditions are imposed at both radial ends.

On the other hand,  the frequency-domain Green function for the reflected potential satisfies
\begin{equation}
\left[\frac{\dd^2}{\dd\rstar^2}+\omega^2-V(2C-\rstar)\right]
\widetilde G^{\rm ref}_{\omega}(\rstar,\rstar')
=
\delta(\rstar-\rstar') .
\label{eq4}
\end{equation}
The relation between the original and reflected radial operators implies the exact identity
\begin{equation}
\widetilde G^{\rm ref}_{\omega}(\rstar,\rstar')
=
\widetilde G_{\omega}(2C-\rstar,2C-\rstar') .
\label{eq5}
\end{equation}
and thus, the reflected problem does not introduce a new spectrum. It is the same one-dimensional spectral problem with the two radial ends exchanged.
The same identity also gives the image interpretation of the second distance appearing in the convergence region. If the source point at $\rstar'$ is reflected about $\rstar=C$, its image is at
\begin{equation}
{\rstar'}_{\rm im}
=
2C-\rstar' ,
\label{eq7}
\end{equation}
and hence the distance from the observation point to this image source is 
\begin{equation}
\abs{\rstar-{\rstar'}_{\rm im}}
=
\abs{\rstar+\rstar'-2C} .
\label{eq8}
\end{equation}
Equivalently, in the  centered variable $x$ defined in \eqref{eq:x-rstar}, the image of $x'$ is $-x'$, and the corresponding distance is
$\abs{x-(-x')}
=\abs{x+x'}.
$
This image distance can be described in lightcone language. The radial null coordinates in the two-dimensional $(t,\rstar)$ sector are 
$u=t-\rstar$ and 
$v=t+\rstar$ 
and a  direct radial null ray from $(t',\rstar')$ to $(t,\rstar)$ gives the direct lightcone distance
\begin{equation}
L_{\rm dir}
=
\abs{\rstar-\rstar'} .
\label{eq11}
\end{equation}
On the other hand,
a direct radial null ray from the reflected image source $(t',{\rstar'}_{\rm im})$ to the observation point gives instead
\begin{equation}
L_{\rm im}
=
\abs{\rstar-{\rstar'}_{\rm im}}
=
\abs{\rstar+\rstar'-2C} .
\label{eq22}
\end{equation}
Thus the convergence condition \eqref{eq:conv-reg} can be written as
\begin{equation}
t-t'>
\max\{L_{\rm dir},L_{\rm im}\}.
\label{eq33}
\end{equation}
The QNM expansion therefore converges only after both the direct lightcone and the reflected image lightcone have passed the observation point.

It is important that no reflecting boundary has been imposed at $\rstar=C$. The bounce radius is not a wall in the physical exterior geometry. It is the fixed point of a radial involution in the analytically continued Kruskal plane. The reflected lightcone should therefore not be interpreted as an ordinary classical reflected ray in the exterior Schwarzschild spacetime. Rather, the mirror construction explains why the optical length associated with the additional convergence boundary has the image form $\abs{\rstar+\rstar'-2C}$.

\section{The folded radial problem and the origin of the mirror phase}

The folding formalism is a  way of rewriting a one-dimensional problem on the full line as a multi-component problem on a half-line \cite{Wong:1994np,Oshikawa:1996dj, Bachas:2001vj,Bajnok:2004jd,Ohya:2011qu,Weder:2015}. One chooses a point at which to cut the line and then maps the two sides of the cut to the same positive half-line. The fields on the two sides become different components, or channels, of a vector-valued wavefunction. The endpoint of the half-line becomes a vertex, and the conditions at this vertex encode the fact that the original full-line wavefunction was single-valued and smooth across the cut. Thus folding does not introduce a physical wall or a new reflecting boundary condition. It is an exact change of representation.

In the present problem the natural folding point is the fixed point of the radial mirror map. This formulation is useful because it separates same-side propagation from opposite-side propagation. The former gives the ordinary direct radial distance, while the latter gives the distance to the reflected image of the source. In the following we use this folded description to identify the direct and mirror phases which control the radius of convergence of the QNM expansion.

\subsection{Folding the RW equation}
Here,  
the original radial line in the centered coordinate $-\infty<x<\infty$ is split into two sides, $x>0$ and $x<0$. Folding means that both sides are described using a positive coordinate, which we denote in the sequel  by $\rho>0$. The right side is kept as it is, while the left side is reflected onto the same half-line.
For $\rho>0$, we define
\begin{equation}
\psi_R(\rho)=\psi(C+\rho),
\qquad
\psi_L(\rho)=\psi(C-\rho).
\end{equation}
Thus $\psi_R$ describes the original point to the right of $r_{\ast}=C$, while $\psi_L$ describes the original point equally far to the left of $r_{\ast}=C$, and together they define a two-component half-line field,
\begin{equation}
\Psi(\rho)
=
\begin{pmatrix}
\psi_R(\rho)\\
\psi_L(\rho)
\end{pmatrix}.
\end{equation}
This is the $N=2$ version of the standard folding trick used in quantum-graph scattering where a scalar wavefunction on a line cut at one point is rewritten as a two-component wavefunction on a half-line~\cite{Ohya:2011qu}. The folding point becomes a graph vertex, and smoothness of the original wavefunction becomes a set of vertex matching conditions~\cite{Ohya:2011qu,Kostrykin_1999,Weder:2015}.

Because the folded problem has two components, the folded Green function carries channel indices, and 
then the folded Green function is naturally a $2\times 2$ matrix in channel space
\begin{equation}
G^{\rm fold}_{\omega}
=
\begin{pmatrix}
G_{RR} & G_{RL}\\
G_{LR} & G_{LL}
\end{pmatrix}.
\end{equation}
The entries $G_{RR}$ and $G_{LL}$ describe propagation from a channel back to the same channel. These are the diagonal entries of the folded Green function. They correspond to right-to-right or left-to-left propagation, and they carry the direct optical length.
On the other hand,
the entries $G_{RL}$ and $G_{LR}$ describe propagation from one channel to the other channel. These are the off-diagonal entries of the folded Green function. They correspond to right-to-left or left-to-right propagation, and they carry the folded, or mirror, optical length.
In other words, the original  problem on the full line has been rewritten as a two-component problem on the half-line, so the Green function has channel indices. Same channel propagation is diagonal, while opposite channel propagation is off-diagonal.

\begin{figure}[htbp]
  \centering
\includegraphics[width=0.5\textwidth]{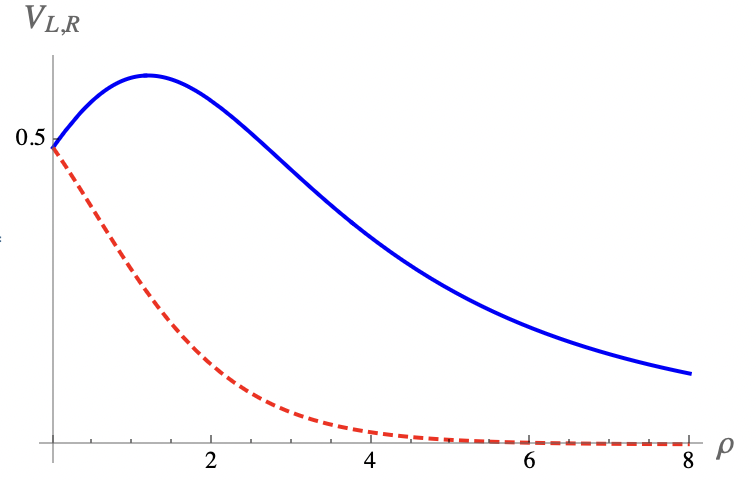}
  \caption{The folded RW potential: The potentials $V_R$ (solid blue curve) and  $V_L$ (dashed red curve)  for the folded problem as a function of $0\leq\rho=\rstar-C$ 
           in units of $2M$ and the particular case $C=0$.}
  \label{fig:VRL}
\end{figure}

The two folded components feel different potentials, see Fig. \ref{fig:VRL},
\begin{equation}
V_R(\rho)=V(C+\rho),
\qquad
V_L(\rho)=V(C-\rho).
\end{equation}
It is important that the folding construction does not assume that the potential is invariant under the radial reflection.  The folded problem is therefore not a symmetric one-channel problem, but a two-channel half-line problem with different channel potentials $V_L(\rho)$ and $V_R(\rho)$.
Thus folding is an exact rewriting of the original full-line radial equation, not a use of a parity symmetry. The reflection symmetry would only be needed if one wanted to identify the two channels. Here the two channels are kept distinct, and the matching conditions at $\rho=0$ simply encode the smoothness of the original wavefunction across the folding point.
Therefore the original full-line radial problem is equivalently rewritten as two half-line equations,
\begin{equation}
\left[
\frac{d^2}{d\rho^2}
+
\omega^2
-
V_R(\rho)
\right]\psi_R(\rho)=0,
\end{equation}
and
\begin{equation}
\left[
\frac{d^2}{d\rho^2}
+
\omega^2
-
V_L(\rho)
\right]\psi_L(\rho)=0.
\end{equation}
The reason this is a useful representation is that the folded picture separates two propagation channels. However, 
the crucial point is what happens at the folding point $\rho=0$. Since the folded fields come from one smooth original wavefunction, they must satisfy
\begin{equation}
\psi_R(0)=\psi_L(0).
\label{eq:match}
\end{equation}
The derivative condition is slightly less obvious because the left side was folded with $C-\rho$. Differentiating $\psi_L(\rho)=\psi(C-\rho)$ gives a minus sign, and therefore smoothness of the original wavefunction demands
\begin{equation}
\frac{d\psi_R}{d\rho}(0)
+
\frac{d\psi_L}{d\rho}(0)
=
0 .
\label{eq:diff-match}
\end{equation}
These two equations \eqref{eq:match} and \eqref{eq:diff-match} are the vertex conditions of the folded problem. They are not boundary conditions for a reflecting wall but they are such that the two half-line components are glued so as to reconstruct the original smooth function on the full line.

Let the observation point and the source point be represented by positive folded coordinates $\rho$ and $\rho'$, respectively. If the source and observer are in the same folded channel, the optical length is the direct one,
$
|\rho-\rho'| .
$
If propagation goes between opposite folded channels, the path length is instead
$
\rho+\rho' .
$
In terms of the original centered tortoise coordinate, this is the difference between a direct path and an image path. Returning to $\rstar$, the direct distance is
$
|\rstar-\rstar'|,
$
whereas the mirror distance is
$
|\rstar+\rstar'-2C| .
$
Thus, the second distance is the distance from the observation point to the reflected image of the source.

At high frequency, the same distinction appears in phases. The diagonal part of the folded Green function, which propagates within the same folded channel, carries phases of the form
\begin{equation}
\exp\left[i\omega |\rho-\rho'|\right].
\end{equation}
The off-diagonal part, which propagates between opposite folded channels, carries phases of the form
\begin{equation}
\exp\left[i\omega(\rho+\rho')\right].
\end{equation}
In the original tortoise coordinate, this off-diagonal phase becomes
\begin{equation}
\exp\left[i\omega(r_{\ast}+r'_{\ast}-2C)\right],
\end{equation}
which is just the mirror phase. The later is simply the phase associated with propagation from one folded channel to the other.

Equivalently, the folded Green function is naturally a two-channel object. Its diagonal part describes propagation within the same folded channel, while its off-diagonal part describes propagation between opposite folded channels. In the WKB approximation, away from turning points, the leading radial propagation is controlled by eikonal phases. In the folded problem there are two natural optical lengths. Propagation within the same folded channel has length $|\rho-\rho'|$, whereas propagation between opposite folded channels has length $\rho+\rho'$. Thus the folded Green function has the schematic large-$|\omega|$ phase structure
\begin{align}
G^{\rm fold}_{\omega}(\rho,\rho')
\sim &
A_{\rm dir}(\omega;\rho,\rho')
\exp\left[\ii \omega|\rho-\rho'|\right]\nonumber \\
+&
A_{\rm mir}(\omega;\rho,\rho')
\exp\left[\ii \omega(\rho+\rho')\right]
+
\cdots .
\end{align}
Here $A_{\rm dir}$ and $A_{\rm mir}$ are not assumed to be constants. They contain the effect of the potential, the Jost transfer matrix, and the vertex matching conditions at the folding point. The important assumption for the convergence problem is that these coefficients do not introduce additional exponential optical lengths in the large-overtone limit.

Near a QNM pole, the pole part of these coefficients gives the residue of the frequency-domain Green function. If the large-overtone residues inherit the two phase sectors displayed above, then one may write 
\begin{align}
{\rm Res}_{\omega=\omega_n}\widetilde G_{\omega}
\supset &
B^{\rm dir}_{n}
\exp\left[\pm \ii \omega_n(\rstar-\rstar')\right]\nonumber \\
+&
B^{\rm mir}_{n}
\exp\left[\pm \ii \omega_n(\rstar+\rstar'-2C)\right].
\label{eq:Green-B-B}
\end{align}
The coefficients 
$B^{\rm dir}_{n}$ and $B^{\rm mir}_{n}$ are the corresponding residue prefactors. The folding argument identifies the two possible optical phases, whereas the further dynamical assumption is that the high-damping residue prefactors are subexponential in the overtone number $n$. 
Using now the large-damping QNM spacing~\cite{Leaver1986,Motl:2003cd,BertiCardosoStarinets2009},
\begin{equation}
\omega_n=-\ii\frac{2\pi}{\beta} n+\mathcal{O}(1),\qquad \beta=8\pi M,
\label{eq:asympt-omega}
\end{equation}
we can determine the convergence condition follows. A typical QNM term contains the time factor together with one of the radial phases,
\begin{equation}
\exp[-i\omega_n(t-t')]
\exp[\pm i\omega_n L]
=
\exp[-i\omega_n(t-t'\mp L)] .
\end{equation}
For the direct channel we have
\begin{equation}
L=L_{\rm dir}=r_{\ast}-\rstar',
\end{equation}
whereas for the mirror channel,
\begin{equation}
L=L_{\rm mir}=\rstar+\rstar'-2C .
\end{equation}
Using the asymptotic form Eq. \eqref{eq:asympt-omega} of $\omega_n$, the magnitude of the exponential part behaves as
\begin{equation}
\left|
\exp[-i\omega_n(t-t'\mp L)]
\right|
\sim
\exp[-\frac{2\pi}{\beta} n(t-t'\mp L)] .
\end{equation}
Cauchy's root test therefore requires
\begin{equation}
t-t'-L>0 \quad \mbox{and}\quad t-t'+L>0 ,
\end{equation}
so that 
\begin{equation}
t-t'>|L| .
\end{equation}
Applying this first to the direct channel and then to the mirror channel gives
\begin{equation}
t-t'>
\max\left\{
|\rstar-\rstar'|,
|\rstar+r'_{\ast}-2C|
\right\}.
\end{equation}
Recapitulating, the original radial line is folded at $\rstar=C$. The two sides become two channels on a half-line. Smoothness of the original wavefunction becomes matching conditions at the folded vertex. The diagonal channel gives the ordinary direct distance, whereas the off-diagonal channel gives the image distance. In other words, the folded formalism naturally produces the mirror lightcone.

\section{Relation to the bouncing-geodesic picture}

There is a simple way to understand the appearence  of the two optical lengths  from the AdS$_2$ Green function. At large tortoise radius, the Schwarzschild radial equation reduces to the equation of a massive scalar on AdS$_2$, with the angular momentum playing the role of the AdS$_2$ mass as has been advocated  in \cite{Kehagias:2025tqi}. Indeed, the Green function for perturbations of  the Schwarzschild background  at $\rstar\gg r_s$ satisfies 
\begin{eqnarray}
\left(\!\frac{\partial^2}{\partial r_*^2}
\!-\!\frac{\partial^2}{\partial t^2}\!-\! \frac{\ell(\ell\!+\!1)}{r_*^2}\!\right)G(t,r_*;t',r_*')\!=\!
\frac{\delta(t\!-\!t')\delta(r_*\!\!-\!r_*')}{\sqrt{-g}}.
\label{green0}
\end{eqnarray}
This equation is identical to the  the equation for  the Green function $G_{\rm AdS_2}$ of a massive scalar with mass square $\ell(\ell+1)$ on  AdS$_2$$\times$S$^2$  as it  has been recognized  in \cite{Kehagias:2025tqi}.  i.e.,  
\begin{eqnarray}
\left(\Box_{\rm AdS_2} -\ell(\ell+1)\right)G_{\rm AdS_2}=
\frac{1}{\sqrt{-g}}\delta(t-t')\delta(r_*-r_*').
\label{green}
\end{eqnarray}
The 
AdS$_2$$\times$S$^2$ spacetime has metric in Poincar\'e coordinates for the AdS$_2$ 
\begin{eqnarray}
\dd s^2=\frac{-\dd t^2+\dd r_*^2}{r_*^2}+\dd \Omega_2^2,
\label{ads}
\end{eqnarray}
whereas  the boundary is at $\rstar=0$.
 The corresponding Green function is naturally written as \cite{Avis:1977yn,Burgess:1984ti,Satoh:2002bc}
\begin{align}
G_{\rm AdS_2}(\chi)=&\frac{1}{2^\ell\sqrt{\pi}}
\frac{\Gamma(\ell+1)}{\Gamma(\ell+\frac{3}{2})} 
\chi^{-\ell-1}\,
{}_2F_1\left(\frac{\ell}{2}+\frac{1}{2},\frac{\ell}{2}+1,\ell+\frac{3}{2},\frac{1}{\chi^2}\right), 
\label{g1}
\end{align}
where $\chi$ is the AdS$_2$ invariant distance defined  as 
\begin{equation}
\chi
=
\frac{-(t-t')^2+\rstar^2+{\rstar'}^{2}}
{2\rstar \rstar'} .
\end{equation}
Thus,  the Green function is proportional to a hypergeometric function whose argument is $z=\chi^2$. Since the hypergeometric function has branch points at  $z=1$ and $z=\infty$, the radius of convergence
of the hypergeometric series is $|z|=1$, which in the $\chi$-variable is $|\chi|>1$, giving   the boundary of the convergence domain 
\begin{equation}
\chi^2=1 .
\end{equation}
Therefore, the two possibilities are $\chi=1$ and $\chi=-1$. The first gives
\begin{equation}
-(t-t')^2+\rstar^2+{\rstar'}^{2}
=
2\rstar \rstar',
\end{equation}
or equivalently
\begin{equation}
(t-t')^2
=
(\rstar-{\rstar'})^2 ,
\end{equation}
which corresponds to  the ordinary direct lightcone. The second possibility gives
\begin{equation}
-(t-t')^2+\rstar^2+{\rstar'}^{2}
=
-2\rstar \rstar',
\end{equation}
or 
\begin{equation}
(t-t')^2
=
(\rstar+\rstar')^2 .
\end{equation}
which  is the image, or reflected, lightcone. In the retarded late-time region, where $t-t'>0$, the hypergeometric convergence condition therefore becomes
\begin{equation}
t-t'
>
\max\left\{
|\rstar-\rstar'|,
|\rstar+\rstar'|
\right\}.
\end{equation}
Thus the radius of convergence obtained from the AdS$_2$ hypergeometric Green function is precisely the same as the one obtained from the folded radial formalism.

The geometric interpretation is also transparent. In Poincare AdS$_2$, radial null geodesics are unaffected by the conformal factor in the metric, so that they are identical to the null geodesics of flat spacetime, i.e., 
\begin{equation}
t-\rstar={\rm const.}\qquad
\mbox{or}\qquad t+r_{\ast}={\rm const.}\, .
\end{equation}
Clearly, the direct null geodesic from the source to the observer gives the lightcone distance
\begin{equation}
L_{\rm dir}=|r_{\ast}-{\rstar'}| .
\end{equation}
However, because the Poincare patch has a boundary at the origin of the radial coordinate $\rstar=0$, there is also an image trajectory. This trajectory goes from the source to the AdS$_2$ boundary and then from the boundary to the observer. Its total optical length is $
\rstar+{\rstar'}$, 
which is, equivalently, the direct lightcone distance from the image source at $-{\rstar'}$ to the observer at $\rstar$
\begin{equation}
L_{\rm mir}=|r_{\ast}-(-{\rstar'})|
=
|\rstar+{\rstar'}| ,
\end{equation}
matching  exactly the mirror distance found in the folded description.

The folded and AdS$_2$ pictures above  identify the two
optical lengths that bound the convergence region, but they leave open the relative strength with which the direct and mirror lightcones enter the
Green function. This is precisely the dynamical input in Eq. \eqref{eq:Green-B-B}, where one
must assume that the high-overtone residue prefactors $B^{\rm dir}_n$ and
$B^{\rm mir}_n$ are subexponential in the overtone number, so that the mirror
optical length is not shifted. We will show below that the AdS$_2$ Green function \eqref{g1} determines this relative weight exactly, verifying  the
assumption.

The Green function in Eq. \eqref{g1} involves the zero-balanced hypergeometric function
${}_2F_1\left(\frac{\ell}{2}+\frac{1}{2},\frac{\ell}{2}+1,\ell+\frac{3}{2},\frac{1}{\chi^2}\right)$. The latter diverges logarithmically at  $\chi^2=1$ since 
\begin{equation}
  {}_2F_1(a,b;a+b;z)
  \;\overset{z\to1}{=}\;
  -\frac{\Gamma(a+b)}{\Gamma(a)\Gamma(b)}
  \Big[\ln(1-z)+\psi(a)+\psi(b)+2\gamma_E\Big]
  +\mathcal{O}\!\big((1\!-\!z)\ln(1\!-\!z)\big)
  \label{eq:hyperlog}
\end{equation}
for zero-balanced hypergeometric functions $(c=a+b)$.
Therefore, the Green function in Eq. \eqref{g1} has logarithic diverges 
\begin{align}
  G_{\mathrm{AdS}_2}(\chi)\;\overset{\chi\to\pm1}{\simeq}\;&
  \chi^{-\ell-1}\big[\ln(\chi-1)+\ln(\chi+1)\big]+(\text{regular})\nonumber \\
\;\overset{\phantom{\chi\to\pm1}}{\simeq}\;&
  |\chi|^{-\ell-1}\big[\ln(\chi-1)+(-1)^{-\ell-1}\ln(\chi+1)\big]+(\text{regular}),
  \label{eq:Green-log}
\end{align}
where we have used the fact that the branch of $\chi^{-\ell-1}$
at $\chi=-1$  gives
\begin{eqnarray}
\chi^{-\ell-1}\to e^{\mp i\pi(\ell+1)}\,|\chi|^{-\ell-1}=(-1)^{\ell+1}\,
|\chi|^{-\Delta}.
\end{eqnarray}  
Then using 
\begin{equation}
\chi\mp1=\frac{(L_--(t-t'))(L_++(t-t'))}{2\rstar\rstar'}\;\overset{\chi\to\pm1}{\simeq}\;\frac{L_\pm}{r_*r'_*}\,\big(L_\pm-(t-t')\big),
  \label{eq:linearize}
\end{equation}
 where 
\begin{eqnarray}
  L_-=L_{\rm dir}=|r_*-r'_*|,\qquad L_+=L_{\rm mir}=r_*+r'_*,
\end{eqnarray}
we find that the $g_n(r,r')$ of Eq. (\ref{eq:qnm-sum}) is 
\begin{equation}
g_n(r,r')\;\simeq\;
\frac{1}{\pi \,n}
  \Big[\,e^{\,i\omega_n|r_*-r'_*|}
  \;+\;(-1)^{\ell+1}\,e^{\,i\omega_n(r_*+r'_*-2C)}\,\Big].
  \label{eq:residue}
\end{equation}
Notice that at the full problem,  there will be  $2M/r$ and $1/\ell$ corrections to the leading $(-1)^{\ell+1}$ contribution of the mirror source at $L_{\rm mir}=\rstar+\rstar'$.

This AdS$_2$ picture above also clarifies the relation to bouncing geodesics. In AdS$_2$, the bounce is a genuine null ray that reaches the timelike boundary and returns to the bulk. The hypergeometric Green function encodes both the direct and boundary-reflected lightcones through its singular surfaces $\chi=\pm1$ which is the same structure that appears as diagonal and off-diagonal propagation in the folded problem.
The Schwarzschild case is similar but has no physical reflecting wall. There, the second convergence boundary arises from a complex-time bouncing singularity \cite{ArnaudoWithers2026} rather than a real reflected ray leading to two complementary viewpoints. The bouncing-geodesic analysis, which is  a spacetime argument, locates the relevant singularity of the retarded Green function directly, while our mirror construction, which is a radial-scattering argument, explains why its optical length takes the form of a distance to a reflected source. In this picture the bounce radius is the fixed point of the radial involution, and the reflected lightcone is the real radial image of the complex-time singularity, not a ray bouncing off a wall.
This is  important since   isospectrality explains why the mirror problem shares the original QNM frequencies, but it does not  fix the convergence region, which also depends on the large-overtone residues. The required dynamical input is that the image optical phase appears in those residues with no extra exponential factor so that the same high-damping information carried by the asymptotic QNM and the residue structure \cite{Leaver1986,Motl:2003cd,BertiCardosoStarinets2009}.

\section{Conclusion}

We have argued that the Schwarzschild QNM convergence region admits a simple radial mirror interpretation. The key observation is that the tortoise coordinate allows a natural reflection about a distinguished radial point. This reflection does not define a symmetry of the full Schwarzschild spacetime, nor does it leave the RW potential invariant. Instead, it maps the original radial problem to a reflected radial problem, where the two radial operators are related by conjugation and therefore are isospectral, provided the boundary conditions are reflected at the same time.

This mirror map gives a geometric interpretation of the additional lightcone distance that appears in the convergence region. Besides the ordinary direct lightcone between source and observer, there is an image lightcone obtained by reflecting the source about the fixed point of the radial involution. In this sense, the second convergence boundary is the lightcone from the mirror image of the source.

The folded formulation makes this interpretation especially transparent. After folding the radial line at the reflection point, the original one-dimensional scattering problem becomes a two-channel half-line problem. Propagation within the same folded channel gives the direct contribution, while propagation between opposite folded channels gives the mirror contribution. The mirror phase is therefore not imposed by hand. It appears as the off-diagonal propagation phase of the folded radial problem.

The construction should be understood as a radial scattering interpretation, not as an independent proof based only on isospectrality. The convergence of the QNM expansion also depends on the large-overtone residues of the original Schwarzschild Green function. Under the standard high-damping assumption that the relevant residue prefactors do not introduce additional exponential lengths, the direct and mirror optical lengths determine the convergence boundary.

The bounce radius is then identified with the fixed point of the radial mirror map. It is not the horizon, not the light ring, and not the maximum of the RW potential. Rather, it is selected by the global analytic structure of the radial problem and by the reflected image of the Schwarzschild singularity in the radial Kruskal plane. This explains why the same point appears naturally in the bouncing-singularity picture, while giving it a complementary interpretation in terms of mirror scattering and folded radial propagation.

\section*{Acknowledgements}
  A.R.  acknowledges support from the Swiss National Science Foundation (project number CRSII5\_213497). 

\bibliographystyle{JHEP}
\bibliography{mirror-bounce}

@article{Kuntz:2026xep,
    author = "Kuntz, Adrien and Della Rocca, Matteo",
    title = "{Vanishing of all redshift modes in Schwarzschild ringdown}",
    eprint = "2606.07173",
    archivePrefix = "arXiv",
    primaryClass = "gr-qc",
    month = "6",
    year = "2026"
}

@article{DeAmicis:2026wqd,
    author = "De Amicis, Marina and Cannizzaro, Enrico and Carullo, Gregorio and Kuntz, Adrien and Sberna, Laura",
    title = "{Dynamical quasinormal mode excitation II: propagation and convergence in Schwarzschild}",
    eprint = "2605.16492",
    archivePrefix = "arXiv",
    primaryClass = "gr-qc",
    month = "5",
    year = "2026"
}

@article{Motl:2003cd,
  author        = "Motl, Lubos and Neitzke, Andrew",
  title         = "{Asymptotic black hole quasinormal frequencies}",
  journal       = "Adv. Theor. Math. Phys.",
  volume        = "7",
  pages         = "307--330",
  year          = "2003",
  eprint        = "hep-th/0301173",
  archivePrefix = "arXiv"
}

@article{Ohya:2011qu,
    author = "Ohya, Satoshi",
    title = "{Path Integral on Star Graph}",
    eprint = "1104.5481",
    archivePrefix = "arXiv",
    primaryClass = "hep-th",
    reportNumber = "IFUP-TH-2011-9, IFUP-TH/2011-9",
    doi = "10.1016/j.aop.2012.02.009",
    journal = "Annals Phys.",
    volume = "327",
    pages = "1668--1681",
    year = "2012"
}

@article{Kostrykin_1999,
   title={Kirchhoff’s rule for quantum wires},
   volume={32},
   ISSN={1361-6447},
   url={http://dx.doi.org/10.1088/0305-4470/32/4/006},
   DOI={10.1088/0305-4470/32/4/006},
   number={4},
   journal={Journal of Physics A: Mathematical and General},
   publisher={IOP Publishing},
   author={Kostrykin, V and Schrader, R},
   year={1999},
   month=Jan, pages={595–630} }

@article{Weder:2015,
  author = {Weder, Ricardo},
  title = {Scattering Theory for the matrix Schr\"odinger operator on the half line with general boundary conditions},
  journal = {J. Math. Phys.},
  volume = {56},
  pages = {092102},
  year = {2015},
  eprint = {1505.01879},
  archivePrefix = {arXiv},
  primaryClass = {math-ph}
}

@article{ArnaudoWithers2026,
author        = {Arnaudo, Paolo and Withers, Benjamin},
title         = {{Bouncing singularities in Schwarzschild: a geometric origin of the QNM convergence region}},
eprint        = {2605.16489},
archivePrefix = {arXiv},
primaryClass  = {gr-qc},
year          = {2026}
}

@article{Arnaudo:2025uos,
    author = "Arnaudo, Paolo and Carballo, Javier and Withers, Benjamin",
    title = "{Beyond quasinormal modes: a complete mode decomposition of black hole perturbations}",
    eprint = "2510.18956",
    archivePrefix = "arXiv",
    primaryClass = "gr-qc",
    month = "10",
    year = "2025"
}

@article{Arnaudo:2025kit,
    author = "Arnaudo, Paolo and Withers, Benjamin",
    title = "{Price's law from quasinormal modes}",
    eprint = "2511.17703",
    archivePrefix = "arXiv",
    primaryClass = "gr-qc",
    month = "11",
    year = "2025"
}

@article{Leaver1986,
author  = {Leaver, Edward W.},
title   = {{Spectral decomposition of the perturbation response of the Schwarzschild geometry}},
journal = {Phys. Rev. D},
volume  = {34},
pages   = {384--408},
year    = {1986},
doi     = {10.1103/PhysRevD.34.384}
}

@article{BertiCardosoStarinets2009,
author        = {Berti, Emanuele and Cardoso, Vitor and Starinets, Andrei O.},
title         = {{Quasinormal modes of black holes and black branes}},
journal       = {Class. Quant. Grav.},
volume        = {26},
pages         = {163001},
year          = {2009},
eprint        = {0905.2975},
archivePrefix = {arXiv},
primaryClass  = {gr-qc},
doi           = {10.1088/0264-9381/26/16/163001}
}

@article{Oshikawa:1996dj,
    author = "Oshikawa, Masaki and Affleck, Ian",
    title = "{Boundary conformal field theory approach to the critical two-dimensional Ising model with a defect line}",
    eprint = "cond-mat/9612187",
    archivePrefix = "arXiv",
    doi = "10.1016/S0550-3213(97)00219-8",
    journal = "Nucl. Phys. B",
    volume = "495",
    pages = "533--582",
    year = "1997"
}

@article{Bachas:2001vj,
    author = "Bachas, C. and de Boer, J. and Dijkgraaf, R. and Ooguri, H.",
    title = "{Permeable conformal walls and holography}",
    eprint = "hep-th/0111210",
    archivePrefix = "arXiv",
    reportNumber = "CALT-68-2361, CITUSC-01-045, ITFA-2001-33, LPTENS-01-42",
    doi = "10.1088/1126-6708/2002/06/027",
    journal = "JHEP",
    volume = "06",
    pages = "027",
    year = "2002"
}

@article{Bajnok:2004jd,
    author = "Bajnok, Z. and George, A.",
    title = "{From defects to boundaries}",
    eprint = "hep-th/0404199",
    archivePrefix = "arXiv",
    doi = "10.1142/S0217751X06025262",
    journal = "Int. J. Mod. Phys. A",
    volume = "21",
    pages = "1063--1078",
    year = "2006"
}

@article{Wong:1994np,
    author = "Wong, E. and Affleck, I.",
    title = "{Tunneling in quum wires: A Boundary conformal field theory approach}",
    eprint = "cond-mat/9311040",
    archivePrefix = "arXiv",
    doi = "10.1016/0550-3213(94)90479-0",
    journal = "Nucl. Phys. B",
    volume = "417",
    pages = "403--438",
    year = "1994"
}

@article{Kehagias:2025tqi,
    author = "Kehagias, Alex and Riotto, Antonio",
    title = "{AdS perspective on the nonlinear tails in black hole ringdowns}",
    eprint = "2506.14475",
    archivePrefix = "arXiv",
    primaryClass = "gr-qc",
    doi = "10.1103/yk6g-sdbw",
    journal = "Phys. Rev. D",
    volume = "112",
    number = "8",
    pages = "084068",
    year = "2025"
}

@article{Burgess:1984ti,
    author = "Burgess, C. P. and Lutken, C. A.",
    title = "{Propagators and Effective Potentials in Anti-de Sitter Space}",
    reportNumber = "UTTG-29-84",
    doi = "10.1016/0370-2693(85)91415-7",
    journal = "Phys. Lett. B",
    volume = "153",
    pages = "137--141",
    year = "1985"
}

@article{Satoh:2002bc,
    author = "Satoh, Yuji and Troost, Jan",
    title = "{On time dependent AdS / CFT}",
    eprint = "hep-th/0212089",
    archivePrefix = "arXiv",
    reportNumber = "MIT-CTP-3330, UTHEP-466",
    doi = "10.1088/1126-6708/2003/01/027",
    journal = "JHEP",
    volume = "01",
    pages = "027",
    year = "2003"
}

@article{Avis:1977yn,
    author = "Avis, S. J. and Isham, C. J. and Storey, D.",
    title = "{Quantum Field Theory in anti-De Sitter Space-Time}",
    reportNumber = "ICTP-77-78-4",
    doi = "10.1103/PhysRevD.18.3565",
    journal = "Phys. Rev. D",
    volume = "18",
    pages = "3565",
    year = "1978"
}

\end{document}